# Mapping Orthorhombic Domains with Geometrical Phase Analysis in Rare-Earth Nickelate Heterostructures

*Bernat Mundet*[1,2,*,¥]*, Marios Hadjimichael*[1]*, Jennifer Fowlie*[1,¶]*, Lukas Korosec*[1]*, Lucia Varbaro*[1]*, Claribel Dominguez*[1]*, Jean-Marc Triscone*[1] *and Duncan T. L. Alexander*[2]

[1]Department of Quantum Matter Physics, University of Geneva, Geneva, Switzerland.

[2]Electron Spectrometry and Microscopy Laboratory (LSME), Institute of Physics (IPHYS), École Polytechnique Fédérale de Lausanne (EPFL), Lausanne, Switzerland.

ABSTRACT

Most perovskite oxides belong to the *Pbnm* space group, composed by an anisotropic unit cell, *A*-site antipolar displacements and oxygen octahedral tilts. Mapping the orientation of the orthorhombic unit cell in epitaxial heterostructures that consist of at least one *Pbnm* compound is often required to understand and control the different degrees of coupling established at their coherent interfaces and, therefore, their resulting physical properties. However, retrieving this information from the strain maps generated with high-resolution scanning transmission electron microscopy can be challenging, because the three pseudocubic lattice parameters are very similar in these systems. Here, we present a novel methodology for mapping the crystallographic orientation in *Pbnm* systems. It makes use of the geometrical phase analysis algorithm, as applied to aberration-corrected scanning transition electron microscopy images, but in an unconventional way. The method is fast and robust, giving real-space maps of the lattice orientations in *Pbnm* systems, from both cross-sectional and plan-view geometries and across large fields of view. As an example, we apply our methodology to rare-earth nickelate heterostructures, in order to investigate how the crystallographic orientation of these films depends on various structural constraints that are imposed by the underlying single crystal substrates. We observe that the resulting domain distributions and associated defect landscapes mainly depend on a competition between the epitaxial compressive/tensile and shear strains, together with the matching of atomic displacements at the substrate/film interface. The results point towards strategies for controlling these characteristics by appropriate substrate choice.

TEXT

**I. Introduction**

Transition metal perovskite oxides, with the chemical formula $AB\text{O}_3$, consist of a pseudocubic (PC) structure composed of a central *B*-site transition metal cation octahedrally coordinated with 6 oxygen atoms (face centered positions) and eight *A*-site cations situated at the corner-sharing positions. In the ideal cubic structure, the *B*—O—*B* bond angle between the transition metal and the neighboring oxygen atoms is 180º. However, depending on the relative sizes of the *A*-site and *B*-site cations, the system may lower its symmetry, adopting additional atomic displacements and/or octahedral rotations,[1] which may ultimately modify this bond angle. Common examples of distorted perovskite lattices are the orthorhombic *Pbnm* (or *Pnma*) lattice – characterized by in-phase octahedral tilts about the longest orthorhombic axis ($c_{ORT}$) and out-of-phase tilts about the other two pseudocubic (PC) axes – and the rhombohedral $R\bar{3}c$ lattice, with out-of-phase rotations about all three PC axes. These two structures are respectively described by the following octahedral tilt patterns in Glazer notation: $a^-a^-c^+$ and $a^-a^-a^-$.[2–4] In conjunction, the *B*—O—*B* bond angle becomes modified. The physical properties of these oxides are in turn influenced by the geometry of this bond, as it regulates the orbital overlapping between the *2p* and *d* electronic levels of the oxygen and transition metal respectively.[5–8] Tuning this structural parameter is therefore a common strategy for tailoring their physical properties.[9–12] This is often done by growing epitaxial heterostructures, where the bond characteristics are modified due to a biaxial strain imposed by the lattice mismatch from the underlying single crystal substrate.[13–15] The characteristic length-scale associated with this strain effect can be rather long (of the order of tens of nanometers), after which it may be partially relieved through the incorporation of lattice defects in the film.[16,17] Over a shorter length-scale range (a few unit cells), this bond angle can be more drastically modified by engineering an epitaxial and coherent interface between two perovskite materials with dissimilar *B*—O—*B* bond angles.[10,18–20] In this scenario, however, other interfacial phenomena such as charge transfer, polar discontinuities, and orbital reconstructions, among others, may also affect the behavior of the heterostructures, and therefore need to be considered to understand the physical properties of the engineered systems.[21–25] In addition to this, a symmetry mismatch between two neighboring epitaxial oxide layers may also affect the structural properties of the system, spanning over a length-scale intermediate to the two aforementioned contributions.[26–31] An example of this occurs in compounds with *Pbnm* symmetry: the constraint imposed on the deposited film depends on the orientation of the lattice with respect to the underlying substrate. For instance, when the substrate is cubic, the unit cell is shear strained when the long orthorhombic axis ($c_{ORT}$) is oriented perpendicular to the interface with the substrate ("out-of-plane"), whereas this shear strain does not develop when $c_{ORT}$ lies in the substrate plane ("in-plane").[32,33] As the resulting physical properties of the deposited films may depend on their precise crystallographic orientation and associated structural domain morphology,[27,28,34–36] identifying and controlling these structural parameters is paramount to properly tailoring their physical properties. For instance, this is particularly relevant for engineering improper ferroelectricity in *Pbnm* systems, where an out-of-plane $c_{ORT}$ axis is required.[37–40] Moreover, depending on the crystallographic orientation, distinct multidomain configurations may be present in the films, which can ultimately affect their defect landscape and physical properties.

The crystallographic orientation of *Pbnm* systems can be assessed from the characteristic half-order reflections appearing in X-ray diffraction (XRD) or electron diffractograms.[41–43] However, these techniques do not provide real-space information about their domain characteristics. For this purpose, here we describe an efficient methodology to map the distribution of *Pbnm* domains and associated boundaries in real space, using scanning transmission electron microscopy (STEM). Our technique is based on the geometrical phase analysis (GPA) algorithm, as commonly used to map strain fields in high-resolution (S)TEM images.[44] While GPA is known to be useful for mapping polarization textures in ferroelectric samples – thanks to the large difference between the *a*/*c* parameters in tetragonal systems[45,46] – it has not been yet exploited to map *Pbnm* domains, because the three PC lattice parameters are all of similar length in this structure. Despite this fact, here we show that GPA is still suitable to map the distribution of *Pbnm* domains, by applying specific (and unconventional) virtual aperture settings. After its demonstration, we employ our method on a series of orthorhombic (*Pbnm*) rare-earth nickelate heterostructures grown under different epitaxial constraints. This particular oxide family has been widely studied, because it displays a bandwidth-controlled metal-insulator transition (MIT) that is directly regulated by the Ni—O—Ni bond angle.[6,14,47,48] We propose that the orientation of the crystallographic structure and the resulting domain distribution is set by an energy competition between three contributions; two elastic terms related to epitaxial strain (one component from the normal strain, the other from the shear strain), and an interfacial term that depends on the coupling of octahedral rotations and/or atomic displacements at the interfaces. The methodology described in this work, and all the main conclusions, can be directly transferred to any *Pbnm* system, as well as other systems presenting similar structural distortions or other additional reflections in the Fourier transform patterns of their atomic resolution STEM images.

**II. Methodology description**

The *Pbnm* crystal structure of NdNiO$_3$ is displayed in Figure 1, viewed along the $[001]_{ORT}$ (or $c_{ORT}$) zone axis, panel (a), and the $[1\bar{1}0]_{ORT}$ zone axis, panels (b) and (c). This structure presents an $a^-a^-c^+$ tilt pattern, which means that the oxygen octahedra rotate in-phase (+ sign) and out-of-phase (− sign) about one and two PC axes, respectively.

To partially compensate the oxygen octahedral tilts, additional *A*-site (rare-earth, here Nd) displacements appear, which run perpendicular to the $c_{ORT}$ axis direction, as shown by the arrows in Figure 1(b) and 1(c). These *A*-site displacements are commonly referred to as antipolar motion displacements (or X$_5^-$ mode) and are coupled with the octahedral tilts through a trilinear energetic term.[49,50] Depending on the orientation of the *Pbnm* unit cell, these cations can be either displaced vertically or horizontally with respect to the film/surface interface, as in Figure 1(b) and (c) respectively. The octahedral tilts and the *A*-site displacements double the length of the PC unit cell, leading to the appearance of additional Bragg reflections at half-order peak positions, as shown in the Fourier transform (FT) patterns of Figure 1 (pink arrows in the bottom row). Identifying the reciprocal space positions of these specific half-order reflections in XRD or electron diffractograms therefore allows the *Pbnm* crystallographic orientation to be determined.[17,51,52] Moreover, when appropriate aperture settings are chosen (see yellow circles in the FT patterns of Figure 1), the GPA of the STEM images detects these antipolar motion displacements as parallel and narrow fringes in the shear ($e_{xy}$) and rotation ($r_{xy}$) strain maps when the crystal structure is viewed along the $[110]_{ORT}$ zone axis (see bottom panels of Figure 1(b) and 1(c)). (Note that we do not show the rotation maps in the manuscript because they are equivalent to the shear strain ones.) Therefore, the presence and orientation of these fringes can be used to retrieve the orientation of the orthorhombic unit cell at each image region.

We now describe in detail the reasons behind the appearance of these fringes. We consider a deposited film consisting of several domains with distinct crystallographic orientations. When the acquired HAADF image contains at least two of these domains, multiple half-order reflections will be present within its associated FT pattern. A possible strategy to map the spatial distribution of these domains is by generating inverse FT images from each set of half-order reflections (each one linked to a particular orientation).[53] An example of this is shown in Figure 2, where we display several inverse FT images generated using distinct virtual apertures from a high-angle annular dark field (HAADF) STEM image acquired from a $NdNiO_3$ thin film grown on (001)-oriented $LaAlO_3$. Note that, in this case, two distinct domains are present in the image (see insets), each one generating distinct arrangements of half-order reflections in their associated FT patterns that are like those shown in Figure 1(a) and (b) for the equivalent orientations. Figure 2(b) displays a reconstructed image generated by selecting the $(001)_{PC}$ spots in the FT pattern and performing an inverse FT. The resulting image consists of horizontal parallel fringes that relate to the out-of-plane spacing of the lattice. The same process is repeated in Figure 2(c); this time selecting only the $\left(\frac{1}{2}\ 0\ 1\right)$ *and* $\left(\frac{\bar{1}}{2}\ 0\ 1\right)$ sets of reflections. Since we have selected two sets of reflections, the reconstructed image consists of the summation of the two individual patterns generated from each set of reflections. Nevertheless, it is clearly seen that the resulting chessboard-like pattern is only present in the left-side region, which corresponds to a $[1\bar{1}0]_{ORT}$ zone axis and so is responsible for generating the two selected reflections in the FT pattern. This methodology enables us to correlate specific reflections in the Fourier space with its corresponding real-space area, thus allowing us to map the distribution of domains with distinct crystallographic orientations. However, since a similar pattern is obtained when the $c_{ORT}$ axis is oriented in-plane and out-of-plane (not shown here), these two orientations are not discriminated when using this methodology.

The strategy to overcome this limitation is to combine this chessboard pattern with the one generated using the $[100]_{PC}$ set of reflections. For this, we employ the GPA. The spatial resolution of the generated maps depends on the virtual aperture size, which for classical strain mapping should only include one reflection in order to avoid artefacts. In contrast, in order to achieve our goal, we go against this paradigm. Specifically, we center our reciprocal masks on the $(100)_{PC}$ and $(001)_{PC}$ spots and use an aperture radius larger than $(1/(2c_{PC}))$, such that the $\left(\frac{1}{2}\ 0\ 1\right)_{PC}$ or $\left(1\ 0\ \frac{1}{2}\right)_{PC}$ reflections (if present) are included within the aperture. Note that, since we employ GPA to map orthorhombic domains instead of evaluating strain fields, the same contrast features are obtained regardless of the reference area. Therefore, when the substrate is not viewed in the HAADF image, in general we simply use the whole film image as a reference. In order to illustrate this process, in Figure 2(d) we first display the inverse FT image of the NdNiO$_3$ sample generated using a large virtual aperture that selects both (001)$_{PC}$ *and* $(\frac{1}{2}\,0\,1)_{PC}$ reflections. Now, instead of straight lines such as in panel (b), the fringes follow a wave-like pattern in the left-side area. The same pattern is also obtained by summing the filtered images obtained in panel (b) and (c). When we now apply these large virtual aperture settings to the GPA, the wavy features are translated as local (occurring at each PC unit cell) shear ($e_{xy}$) and rotation ($r_{xy}$) deformations with alternating signs along the $c_{ORT}$ direction. As a result, in the regions where the lattice is viewed along the [110]$_{ORT}$ zone axis, the resulting shear and rotation maps consist of narrow and parallel fringes oriented perpendicular to the $c_{ORT}$ axis. This outcome is seen in the left domain of Figure 2 (e), which shows the antipolar motion map made using this method. Note that the fringes appearing in the antipolar motion maps do not correspond to real strain deformations, as they relate to the intrinsic antipolar motion displacements of the *A* cations. Therefore, from now on we will refer to these maps as antipolar motion maps. In accordance with the principle behind our methodology, reducing the aperture radius below $1/(2C_{ORT})$ (as conventional for GPA) makes these fringes disappear, as shown in Figure 2(f). It should be remarked that, even though these fringes are not present in domains viewed along the $c_{ORT}$ zone axis, we can still distinguish the *Pbnm* symmetries from the other perovskite lattices, e.g. cubic, tetragonal, rhombohedral, because this $c_{ORT}$ zone axis still presents some specific and characteristic half-order reflections in the associated FT pattern, as displayed in Figure 1(a). These reflections arise from the projection of the antipolar displacements of the A-site columns. As a result, the atomic columns are elongated along either the [101]$_{PC}$ or the [$\bar{1}$01]$_{PC}$ direction in an alternating pattern, giving rise to additional peaks corresponding to the {½ 0 ½}$_{PC}$ family of planes. In the following, we apply our method to a series of rare-earth nickelate heterostructures in order to unravel correlations between distinct epitaxial constraints and their resulting domain distributions. We conclude that our adaptation of GPA is a fast and robust way of mapping the possible *Pbnm* orientations simultaneously in both cross-sectional and plan-view images, even when the image data are noisy or show moderate scan distortions.

### III. Mapping *Pbnm* domains in NdNiO$_3$ thin films

We now use the methodology described in the previous section to study the distribution of *Pbnm* domains in NdNiO$_3$ thin films grown on (001)$_{PC}$-oriented LaAlO$_3$, (LaAlO$_3$)$_{0.3}$–(Sr$_2$AlTaO$_6$)$_{0.7}$ (LSAT) and SrTiO$_3$, as well as (110)$_{ORT}$-oriented ($\equiv$ (001)$_{PC}$-oriented) NdGaO$_3$ single crystal substrates. All the studied films have an approximate thickness of 15–20 nm. As the different substrates impose varying degrees of epitaxial strain on the NdNiO$_3$ films, in Table 1 we display the calculated strain tensors for NdNiO$_3$ growth on each of them, for the three possible film orientations.

To evaluate the domain distribution in these systems, we have recorded high-resolution HAADF STEM images of the film cross-sections, as shown in the left panels of Figure 3, and then applied our GPA-based approach to generate their corresponding antipolar motion maps as shown in the right-side panels of the same figure. Considering the strain values presented in Table 1, one could expect that compressive strain will favor an in-plane $c_{ORT}$ axis orientation, whereas tensile strain will stabilize the out-of-plane $c_{ORT}$ direction, since in both cases this would minimize the normal strain. First looking at the compressive case of the NdNiO$_3$ / LaAlO$_3$ system, its results are in agreement with expectations. In its antipolar motion map, shown in the right panel of Figure 3 (a), we identify two different kinds of contrast, both corresponding to domains with the $c_{ORT}$ axis oriented in-plane but with this axis either perpendicular (vertical fringes) or parallel (no fringes) to the viewing axis. This outcome is in agreement with previous reports,[32,42] with $c_{ORT}$ of the film orienting in-plane in order to minimize epitaxial strain. The two different lattice orientations that are observed for this $c_{ORT}$ in-plane configuration result from the equivalence of $a_{PC}$ and $b_{PC}$ in the LaAlO$_3$ substrate, such that domains are equally likely to grow with $c_{ORT}$ parallel to $a_{PC}$ or $b_{PC}$ of LaAlO$_3$.

| Film orientation | LaAlO$_3$ | LSAT | SrTiO$_3$ | NdGaO$_3$ (*Pbnm*, $c_{ORT}$ ‖ $b_{PC}$) |
|---|---|---|---|---|
| $c_{ORT}$ ‖ $a_{PC}$ | $e_{xx}$ = -0.47 %  $e_{yy}$ = -0.40 %  $e_{xy}$ = 0 % | $e_{xx}$ = 1.65 %  $e_{yy}$ = 1.73 %  $e_{xy}$ = 0 | $e_{xx}$ = 2.55 %  $e_{yy}$ = 2.63 %  $e_{xy}$ = 0 | $e_{xx}$ = 1.53 %  $e_{yy}$ = 1.22 %  $e_{xy}$ = 0 |
| $c_{ORT}$ ‖ $b_{PC}$ | $e_{xx}$ = -0.40 %  $e_{yy}$ = -0.47 %  $e_{xy}$ 0 % | $e_{xx}$ = 1.73 %  $e_{yy}$ = 1.65 %  $e_{xy}$ = 0 | $e_{xx}$ = 2.63 %  $e_{yy}$ = 2.55 %  $e_{xy}$ = 0 | $e_{xx}$ = 1.45%  $e_{yy}$ = 1.30%  $e_{xy}$ = 0 |
| $c_{ORT}$ ‖ $c_{PC}$ | $e_{xx}$ = -0.47 %  $e_{yy}$ = -0.47 %  $e_{xy}$ = -0.14% | $e_{xx}$ = 1.65 %  $e_{yy}$ = 1.65 %  $e_{xy}$ = -0.07 % | $e_{xx}$ = 2.55 %  $e_{yy}$ = 2.55 %  $e_{xy}$ = -0.07 % | $e_{xx}$ = 1.45 %  $e_{yy}$ = 1.22 %  $e_{xy}$ = -0.07 % |

**Table 1.** Strain tensor components associated to the crystallographic deformation of the NdNiO$_3$ unit cell when grown on LaAlO$_3$, LSAT, SrTiO$_3$ and NdGaO$_3$ for the three possible film orientations. $a_{PC}$ and $b_{PC}$ are parallel to the substrate surface and $c_{PC}$ is perpendicular to it.

While these basic orientations can also be found with XRD, the STEM-GPA analysis allows us to visualize the domain sizes, which are observed to be rather small (from tens to hundreds of nanometers). The domains are separated by abrupt vertical boundaries, that have an approximate width of one PC unit cell, ~0.4 nm. As we will show later using plan-view imaging, these boundaries run along the $[110]_{PC}$ direction. Therefore, in the cross-section, they sometimes appear to be slightly wider, owing to a projection effect as they are consequently angled by 45° to the incident electron beam.

We now turn to the tensile strain scenario, as applied by the other three substrates. First looking at the results acquired on the NdNiO$_3$ / SrTiO$_3$ system in Figure 3(b), in its antipolar motion map we observe only one kind of contrast – horizontal fringes. This corresponds to an out-of-plane $c_{ORT}$ orientation, as expected for the strain state. However, when we repeat the analysis on the NdNiO$_3$ / LSAT heterostructure, as seen Figure 3 (c), the scenario is more complex. Strikingly, despite the imposed tensile strain, all three possible crystallographic orientations are found to co-exist in the same film, with most domains actually having an in-plane $c_{ORT}$ axis. While a similar multidomain configuration has been reported in other *Pbnm* systems,[42] we point out the ease with which our GPA-based technique directly samples the spatial distribution of the domains. To help interpret these results, we further note that the contrast observed in the LSAT substrate area of the antipolar motion map of Figure 3 (c) is similar to that of the background noise (as observed in the vacuum region above the lamella). This occurs because the intensity of the $[100]_{PC}$ reflections almost vanishes in the LSAT substrate since both *A*-site and *B*-site columns present a similar brightness.

While the tensile strain imposed on this film may make the observed mixture of domain orientations appear unlikely, the full set of imposed strain values in Table 1 provides a possible explanation. If $c_{ORT}$ is oriented out-of-plane, a finite $e_{xy}$ shear deformation is imposed. This occurs because $[1\bar{1}0]_{ORT}$ and $[110]_{ORT}$ of the film are now constrained to be parallel to the *a* and *b* axes of the cubic LSAT substrate, i.e. they are orthogonal, which is not the case in the free *Pbnm* crystal structure. Therefore, even if the tensile strain is smaller when the $c_{ORT}$ is oriented out-of-plane, this is counterbalanced by the ability to relax the $e_{xy}$ shear strain by instead orienting $c_{ORT}$ in-plane. Evidentially, this leads to a competition between the growth of either $c_{ORT}$ orientation when the tensile strain is applied by a cubic substrate. Comparison with the results of growth on cubic SrTiO$_3$, which applies a greater tensile strain and for which only $c_{ORT}$ out-of-plane is observed, indicates that the complete epitaxial orientation transition from in-plane to out-of-plane $c_{ORT}$ axis occurs at a non-zero tensile strain value. Moreover, the fact that we see all three possible orientations co-existing in the same film indicates that this transition is not abrupt but gradual, with the nucleation energy for all the three crystallographic orientations being similar. These results point to the importance of considering more than just the applied compressive/tensile strain to *Pbnm* film growth in determining resultant film orientation.

Such additional considerations come to the fore in the last NdNiO$_3$ film studied in this section, which was grown on (110)-oriented NdGaO$_3$, see Figure 3 (d). As can be observed in its antipolar motion map, we only observe one domain in this film, with the vertical fringes indicating a solely in-plane $c_{ORT}$ axis orientation; even though NdGaO$_3$ imposes similar tensile and shear strain values to the LSAT substrate, which demonstrated a mixed domain configuration. The difference here is that the NdGaO$_3$ substrate and NdNiO$_3$ film share a *Pbnm* symmetry; as a result, the substrate imprints its crystallographic orientation onto the NdNiO$_3$ film because of the structural couplings established at the film/substrate interface. This coupling has been observed in other systems where substrate and film similarly present the same *Pbnm* symmetry.[27,31,42]

The mechanism behind such imprinting of substrate symmetry and hence orientation onto the film is understood to be the minimization of local structural mismatch at the interface, in terms of the key orthorhombic distortions of octahedral tilt pattern and *A*-site displacements. Therefore, in addition to the considerations of epitaxial normal and shear strain, an additional term linked to the structural couplings at the interface (interfacial energy) may also need to be considered for understanding film orientation. Interestingly, we could not find any lattice defects in this film, which we attribute to its monodomain nature resulting from the uniform lattice orientation. Note that this differs to the NdNiO$_3$ /SrTiO$_3$ system. While in the latter system we also observe an out-of-plane $c_{ORT}$ axis orientation, the film may contain twin boundaries separating domains with reversed tilt patterns. For example, the crystallographic defect observed in Figure 3 (b), which is indicated with an orange arrow, could be one of these twin boundaries. The zig-zag pattern of the *A*-site cations (antipolar motion displacement also linked to the $a^-a^-c^+$ tilt pattern) is inverted on either side of this boundary, which is precisely what we would expect when $a_{ORT}$ and $b_{ORT}$ are swapped. However, we cannot rule out either the possibility of this boundary being an antiphase boundary, as it would present a similar structural behavior for the *A*-site sublattice. No such twin (or antiphase) boundaries are observed in the NdNiO$_3$ / NdGaO$_3$ system, stemming from its monodomain configuration that is driven by the substrate symmetry. In comparison, the cubic substrate of the NdNiO$_3$ / SrTiO$_3$ system has no such influence, such that the film can nucleate domains with $a_{ORT}$ parallel to either in-plane unit cell diagonal of the substrate.

To summarize this section, as illustrated in the chart of Figure 3(e), we observe that the crystallographic orientation of *Pbnm* films grown on substrates with cubic cation sublattices (i.e. LaAlO$_3$, LSAT, SrTiO$_3$) is determined by an energetic competition between in-plane normal and shear strain. Under compressive strain, the $c_{ORT}$ axis is oriented in plane, and it changes to an out-of-plane orientation as tensile strain is increased beyond a non-zero value, with both orientations coexisting near the orientation transition. Usefully, our GPA-based approach readily identifies the location of domain boundaries, even when they are difficult to detect visually in the associated HAADF images, such as the apparent twin boundary in Figure 3(b). The different types of domain boundaries may in turn present novel functional properties arising from the structural couplings that need to be established to connect the octahedral tilt patterns between the neighboring domains.

**IV. Domain distribution in $Nd_{1-x}La_xNiO_3$ solid-solution thin films**

Given the strong apparent effect of an otherwise subtle competition between tensile and shear strain on the film orientation and domain landscape, it is interesting to study this balance further. However, experimentally it is challenging to gradually modify the applied epitaxial strain by changing substrates. Therefore, we instead investigate this energetic competition by introducing a certain fraction of La onto the *A*-sites of the film, through a series of $Nd_{1-x}La_xNiO_3$ solid-solution thin films deposited on (001)-oriented LSAT single crystal substrates. By progressively increasing the La/Nd ratio, we can gradually change both strain components, all the while maintaining a tensile strain state. Specifically, here we analyze two films having relative La concentrations of $x = 0.1$ and $x = 0.3$ deposited on LSAT substrates, and later compare two other films, both with $x = 0.4$, deposited on LSAT and $NdGaO_3$ substrates respectively. We do not go beyond this La content because, above $x = 0.4$, orthorhombic and rhombohedral phases co-exist in the same film, as previously reported.[54] Figures 4 (a) and (d) present HAADF images of the films with $x = 0.1$ and $x = 0.3$ respectively.. Unlike the previous analyses made using cross-section views, the STEM samples have been prepared in a plan-view geometry, which offers both a larger overview of the domain distributions, while also being free of the stereological limitations of a cross-section sample with regards to the sampling of domain sizes.

As before, we apply our GPA-based approach to map their orthorhombic domain distributions. The antipolar motion maps are presented in panels (b) for $x = 0.1$ and (e) $x = 0.3$. These maps clearly reveal the same contrast features linked to the *A*-site antipolar displacements as before. Given the large field of view of the input images, this is a clear demonstration of the robustness of our methodology. Because of the change in sample geometry, domains with parallel fringes now correspond to $c_{ORT}$ in-plane. No fringes are observed for $c_{ORT}$ out-of-plane, because the $c_{ORT}$ axis is now parallel to the viewing axis. Usefully, however, as shown in panel (g) the residual contrast for the $c_{ORT}$ out-of-plane domains is different to that of the LSAT substrate, whose similar atomic numbers for the *A*- and *B*-sites give it a different color on the antipolar motion map, as described previously. This allows the two to be discriminated in the antipolar motion maps. The presence of isolated regions of LSAT substrate in the images likely relates to damage induced by the ion milling process during the STEM specimen preparation, as it may entirely remove the film in these areas because of the film's low thickness (15–20nm).

To facilitate the interpretation of the antipolar motion maps, we have colored the areas belonging to the distinct contrast features with different colors representing *Pbnm* domains of the three possible orientations, or substrate area. The resulting sketches are shown in Figures 4 (c) and (f) for the $x = 0.1$ and $x = 0.3$ cases, respectively. Even though we see co-existence of domains with either in-plane or out-of-plane $c_{ORT}$ axis orientation in both films, a clear evolution of their proportions is identified when comparing them. While most domains present the in-plane $c_{ORT}$ axis (magenta and red colors) in the $x = 0.1$ film (like in the pure NdNiO$_3$ thin film), the opposite situation is found in the $x = 0.3$ film, with most domains being oriented with the $c_{ORT}$ axis out-of-plane (green color). To understand this evolution, we have calculated the strain tensors for pure NdNiO$_3$ – using it as a close proxy for the $x = 0.1$ film – and for an Nd$_{0.7}$La$_{0.3}$NiO$_3$ single crystal constrained to an LSAT (100)$_{PC}$-oriented substrate.

| $c_{ORT}$ orientation | NdNiO$_3$ | Nd$_{0.7}$La$_{0.3}$NiO$_3$ |
|:---:|:---:|:---:|
| $c_{ORT} \parallel a_{PC}$ or $c_{ORT} \parallel b_{PC}$ | $e_{xx} = 1.65\ \%$ $e_{yy} = 1.73\ \%$ $e_{xy} = 0$ | $e_{xx} = 1.38\ \%$ $e_{yy} = 1.52\ \%$ $e_{xy} = 0$ |
| $c_{ORT} \parallel c_{PC}$ | $e_{xx} = 1.65\ \%$ $e_{yy} = 1.65\ \%$ $e_{xy} = -0.07\ \%$ | $e_{xx} = 1.38\ \%$ $e_{yy} = 1.38\ \%$ $e_{xy} = -0.36\ \%$ |

**Table 2.** Strain tensor components associated to the crystallographic deformation of the NdNiO$_3$ and Nd$_{0.7}$La$_{0.3}$NiO$_3$ unit cell, when grown on LSAT with the two basic possible film orientations. (Note that $c_{ORT} \parallel b_{PC}$ gives the same values as $c_{ORT} \parallel b_{PC}$, except with $e_{xx}$ and $e_{yy}$ exchanged.) For Nd$_{0.7}$La$_{0.3}$NiO$_3$, we have used the lattice constants of bulk Nd$_{0.7}$La$_{0.3}$NiO$_3$ single crystals that were measured by L. Medarde et al.[55]

Table 2 displays the resultant $e_{xx}$, $e_{yy}$ and $e_{xy}$ values for epitaxial growth of NdNiO$_3$ and Nd$_{0.7}$La$_{0.3}$NiO$_3$ on LSAT. Comparison of these values implies that, as La concentration increases in a $Nd_{1-x}La_xNiO_3$ solid solution, the tensile strain will decrease while, for the $c_{ORT}\|c_{PC}$ orientation, the shear strain will increase. Given our hypotheses in section III, whether considering applied tensile strain or applied shear strain, one would therefore expect a decrease in the fraction of $c_{ORT}$ axis out-of-plane when $x$ is increased. In contrast, Figure 4 shows that, experimentally, we find the opposite. However, the situation here is a bit more complex because, as we observe all three possible crystallographic orientations in both films, their epitaxial strain is probably close to the crossover point and, hence, the domain distribution may be very sensitive to any structural modulation. Moreover, we believe that the key to understanding the observed and surprising domain evolution is to consider not only the average tensile strain, but the difference in strain when swapping the $c_{ORT}$ axis from in-plane to out-of-plane. While in both cases (necessarily) $e_{xx}$ does not change when going from $c_{ORT}$ axis in-plane to out-of-plane, $e_{yy}$ decreases by 0.08% for the $NdNiO_3$ case but a much larger 0.14% for the $Nd_{0.7}La_{0.3}NiO_3$ case. This is because the shape of the pseudocubic unit cell is almost cubic for bulk $NdNiO_3$, but becomes more distorted for both decreasing and increasing size of the $A$-cation.[6] Note that this does not imply that atomic displacements within the unit cell are minimal for $NdNiO_3$. As the La content increases, the difference between pseudocubic lattice parameters increases and the angle between $[1\bar{1}0]_{ORT}$ and $[110]_{ORT}$ (which is related to the difference between $a_{ORT}$ and $b_{ORT}$) departs farther from 90°. Thus, while the energy cost associated to shear strain favors $c_{ORT}$ in-plane more strongly for $x = 0.3$ than for $x = 0.1$ (($\cong NdNiO_3$)), the cost due to normal strain, counter-intuitively, favors $c_{ORT}$ out-of-plane more strongly for $x = 0.3$ than for $x = 0.1$. Our observation that there are more domains with $c_{ORT}$ out-of-plane in the case of $x = 0.3$ indicates that the latter effect is the stronger of the two. In addition, to predict the domain distribution in each film, we should also consider the interfacial term associated to the structural couplings, as the amplitude of the structural distortions linked to the *Pbnm* structure decrease with La content *x*, although this goes beyond the scope of this work.[54]

Beyond giving a detailed overview of the domain distributions, the plan-view images and antipolar motion maps also enable a direct analysis of the domain boundary orientations, which could not be achieved in the cross-section data given the projection effect on these vertical boundaries. This is demonstrated in Figure 5(a), which presents an amplified view of the HAADF image of the $Nd_{0.9}La_{0.1}NiO_3$ film shown in Figure 4 (a) and its corresponding antipolar motion map. It is seen that the boundaries separating domains with in-plane $c_{ORT}$ axis orientation run along the $[110]_{PC}$ direction, as previously stated. In addition to the domain boundaries, lattice defects, specifically Ruddlesden-Popper faults (RPF), are also observed in the data. In the antipolar motion map these faults are seen as straight dark or bright lines, each of which is associated to a structural shift of the lattice by half a PC unit cell along the $[111]_{PC}$ direction. RPFs are commonly observed in rare-earth nickelates.[16,17,56,57] In Figure 5(a), several RPFs are indicated with red arrows; it is seen that they coincide with the domain boundaries.

Given the observed coincidence of RPFs with domain boundaries, it could be suggested that the density of lattice defects will depend on the domain distribution. In order to study this hypothesis, Figures 5 (b) and (c) compare cross-sectional HAADF images and antipolar motion maps from films grown with the same $Nd_{0.6}La_{0.4}NiO_3$ ($x = 0.4$) composition but on two different substrates, LSAT and $NdGaO_3$, respectively. Following the tendency of the $x = 0.1$ and $x = 0.3$ films shown in Figure 4, the film grown on LSAT is primarily composed of domains with $c_{ORT}$ out-of-plane. As indicated by the red arrows in Figure 5(b), many RPFs are again observed at the domain boundaries. A twin (or antiphase) boundary is also seen, as indicated by the orange arrow. This high density of defects contrasts strongly with the data recorded from the $Nd_{0.6}La_{0.4}NiO_3$ film grown on [110]$_{ORT}$-oriented NdGaO$_3$, as shown in Figure 5(c). As for the pure NdNiO$_3$ grown on this substrate shown in Figure 3(d), the film has a monodomain nature with an in-plane *c*$_{ORT}$ axis. Tied to this monodomain nature, it contains no apparent lattice defects. This absence of lattice defects, as compared to the solid-solution film grown on LSAT, is found even though the two films have been grown under the same conditions and have a similar lattice parameter mismatch with the substrate. Therefore, we speculate that the presence of the RPFs in the films grown on LSAT is linked to elastic strain that accumulates at the domain boundaries, which is in turn associated to the anisotropy in the *Pbnm* structure and the structural mismatch (of the octahedral tilts and *A*-site antipolar displacements) between neighboring domains. Our finding therefore shows that the defect landscape of nickelate films can be controlled by choosing an appropriate substrate; one that either minimizes or enhances the density of domains. Such a strategy is of interest because, on the one hand, improving the crystal quality of films by decreasing the density of RPFs is key to achieving and enhancing the superconducting properties of reduced (infinite-layer) nickelate films,[58] while, on the other hand, promoting the presence of RPFs can be of interest in some scenarios, such as improving their electrocatalytic properties.[16,57]

### IV. Summary

In summary, we have presented an efficient tool to map *Pbnm* domains with distinct crystallographic orientations in epitaxial heterostructures, and their corresponding defect landscape, by using an adaptation of GPA. We have exploited our method to map the domain distributions found in a series of rare-earth nickelate heterostructures. This has in turn allowed us to identify that the orientation of the orthorhombic unit cell depends on a competition between in-plane normal and shear strain, and also interfacial energy. Specifically, when the *Pbnm* films are deposited on perovskite substrates having cubic cation sublattices, a compressive epitaxial strain produces an in-plane $c_{ORT}$ axis for the film, consistent with previous reports. In the tensile-strained regime, the film adopts an out-of-plane $c_{ORT}$ axis. However, this orthorhombic orientation also imposes a shear strain on the film, and therefore creates a competition between $c_{ORT}$ out-of-plane and $c_{ORT}$ in-plane. As a result, when the imposed tensile strain is relatively low ($\lesssim 2\%$), a mixture of $c_{ORT}$ out-of-plane and in-plane domains are observed, with their relative proportions apparently controlled by relative tensile strain differences between the two configurations. It is only at high absolute tensile strain ($\gtrsim 2\%$) that a fully $c_{ORT}$ out-of-plane orientation is observed. In comparison, depositing *Pbnm* films on a *Pbnm* substrate introduces another element of control, from an interfacial energy term related to the contiguity of orthorhombic distortions from substrate to film. As a result, despite an applied tensile strain close to 1.5%, for these 15–20 nm thick films a monodomain orientation with $c_{ORT}$ in-plane is created. We have also shown that the density of RPFs depends on the domain distribution, opening new strategies to engineer a desired defect landscape. Further work could be envisaged where these various effects are studied theoretically using first or second principles simulations. Finally, while we have demonstrated our GPA-based methodology on rare earth nickelate thin films, we emphasize that it is a powerful approach suitable for studying the crystallographic orientation for any *Pbnm* system. Given that *Pbnm* is the most common space group for perovskite compounds, it can therefore be widely applied to investigating strain and coupling effects on epitaxial heterostructures, bringing insights that can help in the overall quest for engineering novel systems with enhanced functionalities, ranging from unconventional superconductors to electrochemical catalysts.

*Experimental section:* Films were grown by radiofrequency off-axis magnetron sputtering as described in previous works.[54,59,60] The $Nd_{1-x}La_xNiO_3$ solid solution was achieved by intermittently sputtering from two stoichiometric ceramic targets of $NdNiO_3$ and $LaNiO_3$.

HAADF STEM images were acquired with a double aberration-corrected Thermo Fisher Scientific (FEI) Titan Themis 60-300, operated at 300 kV high tension and using a probe convergence semi-angle of 20mrad and current of 40 pA. To minimize scan distortions in the data, each image was obtained by averaging an acquisition series composed of 20 frames, with consecutive 90º rotations between images. Before averaging, linear and non-linear distortions were corrected using rigid and non-rigid registrations with the SmartAlign plug-in for Digital Micrograph.[61]

AUTHOR INFORMATION

**Corresponding Author**

*Bernat Mundet, bmundet@icn2.cat

**Present Addresses**

¥ Bernat Mundet: Catalan Institute of Nanoscience and Nanotechnology (ICN2), Spain

¶ Jennifer Fowlie: Stanford University, California, USA

**Author Contributions**

J. F., M. H., L.V and C. D. grew the films that were studied in this work. B. M. acquired and processed the experimental data shown in this manuscript. L. K. calculated the strain tensor for all the studied systems. B.M. and D. T. L. A. wrote the manuscript with input from all authors.

All authors participated in the scientific discussions. All authors have given approval to the final version of the manuscript.

ACKNOWLEDGMENT

This work was partly supported by the Swiss National Science Foundation through Division II Grant No. 200020_179155. The research leading to these results received funding from the European Research Council under the European Union's Seventh Framework Program (FP7/2007-2013)/ERC Grant Agreement 319286 Q-MAC). We acknowledge access to the electron microscopy facilities at the Interdisciplinary Centre for Electron Microscopy (CIME), École Polytechnique Fédérale de Lausanne.

ABBREVIATIONS

FT, Fourier transform; GPA, geometrical phase analysis; HAADF, high angle annular dark field; LSAT, $(LaAlO_3)_{0.3}$–$(Sr_2AlTaO_6)_{0.7}$; ORT, orthorhombic; PC, pseudocubic; RPFs, Ruddlesden-Popper Faults; STEM, scanning transmission electron microscopy; TEM, transmission electron microscopy

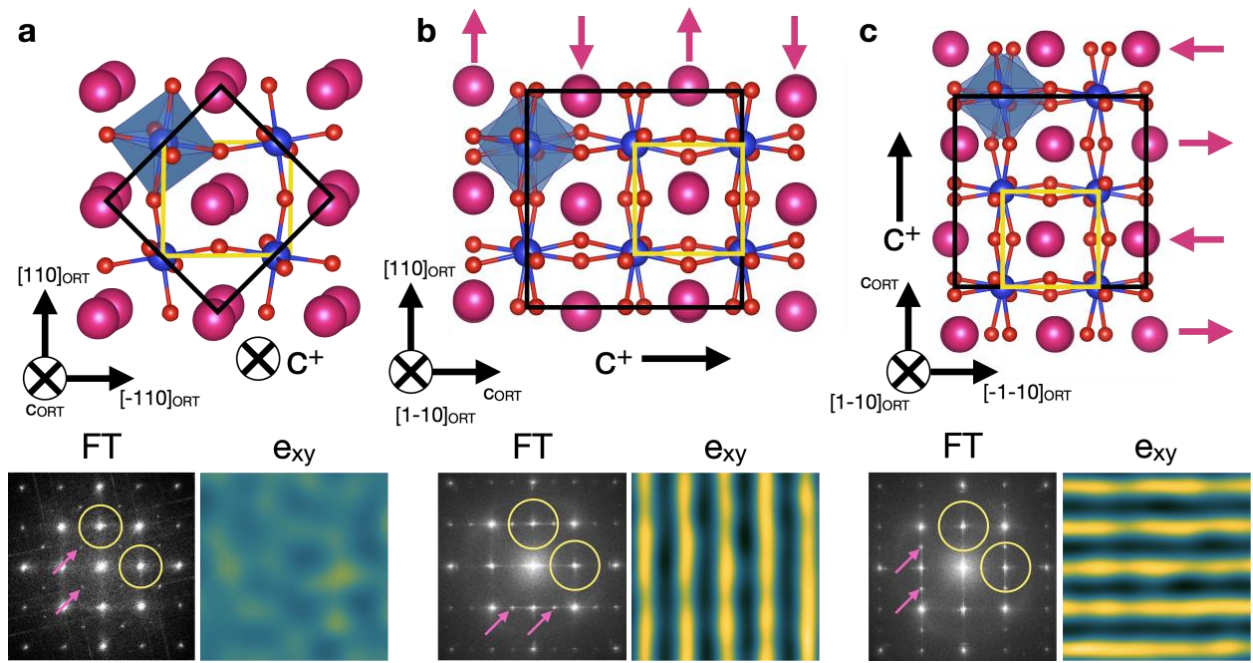

**Figure 1.** *Pbnm* crystal structure viewed along the (a) [001]$_{ORT}$ ($c_{ORT}$) and (b, c) [1$\bar{1}$0]$_{ORT}$ zone axes. The orthorhombic and pseudocubic unit cells are indicated with black and yellow squares, respectively. The magenta arrows indicate the *A*-site antipolar displacements. The lower panels show (left) FT patterns obtained from each corresponding domain type in high-resolution STEM images and (right) the associated features seen in the GPA antipolar motion maps when the yellow circles are used as virtual apertures. The pink arrows in the FT diagrams point to the half-order reflections characteristic of each specific *Pbnm* zone axis and lattice orientation.

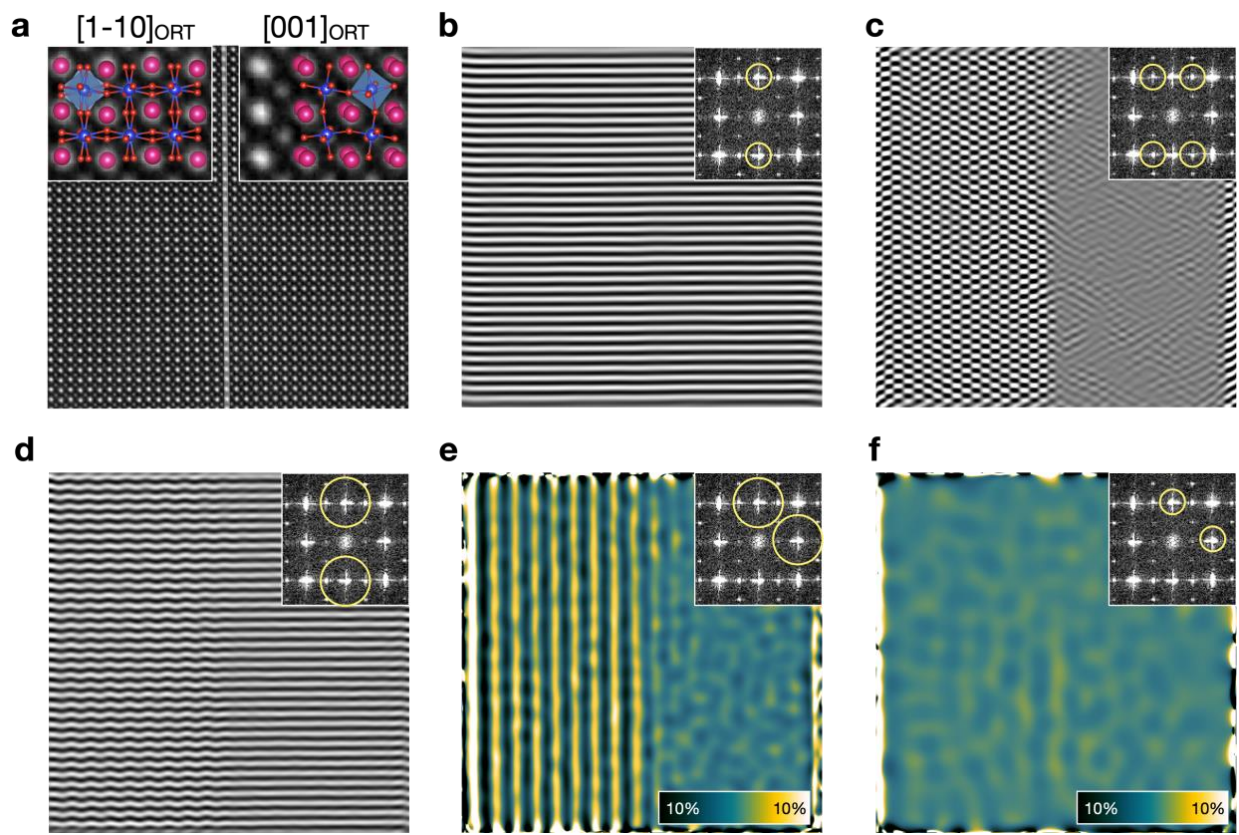

**Figure 2.** (a) Atomic-resolution HAADF STEM image acquired from a NdNiO$_3$ thin film deposited on a (001)$_{PC}$-oriented LaAlO$_3$ single crystal substrate. The insets show a magnified view of the film on either side of the image, with an overlaid illustration of the NdNiO$_3$ unit cell projected along the corresponding zone axis (indicated in the upper label). The boundary separating both domains is indicated with a white line. (b–d) Inverse FT images generated by using different virtual apertures, indicated with yellow circles in the FT patterns that are included in the insets. A wavy pattern is seen in (d) in the left image region, as correlated to the *A*-site AM displacements of its domain. (e-f) antipolar motion maps generated with GPA using different aperture sizes. Narrow vertical fringes appear in the left side of panel (e) due to the wavy pattern observed in panel (d).

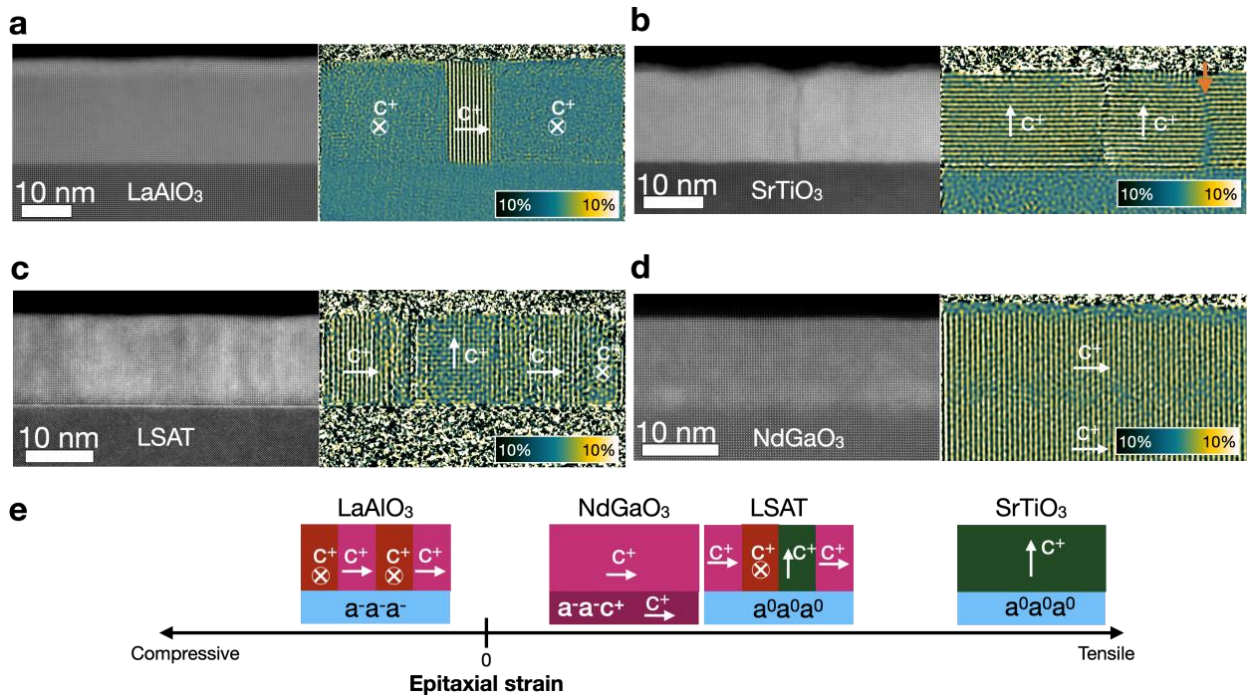

**Figure 3.** HAADF images (left) and associated antipolar motion maps (right) acquired from NdNiO$_3$ films grown on (001)$_{PC}$-oriented (a) LaAlO$_3$, (c) SrTiO$_3$ and (c) LSAT single crystal substrates. The orange arrow points to a twin (or possibly antiphase) boundary. (d) HAADF image (left) and associated antipolar motion map (right) acquired in NdNiO$_3$ film grown on (110)$_{ORT}$-oriented NdGaO$_3$. (e) Illustration showing the evolution of the lattice orientation and associated domain distribution with epitaxial normal strain.

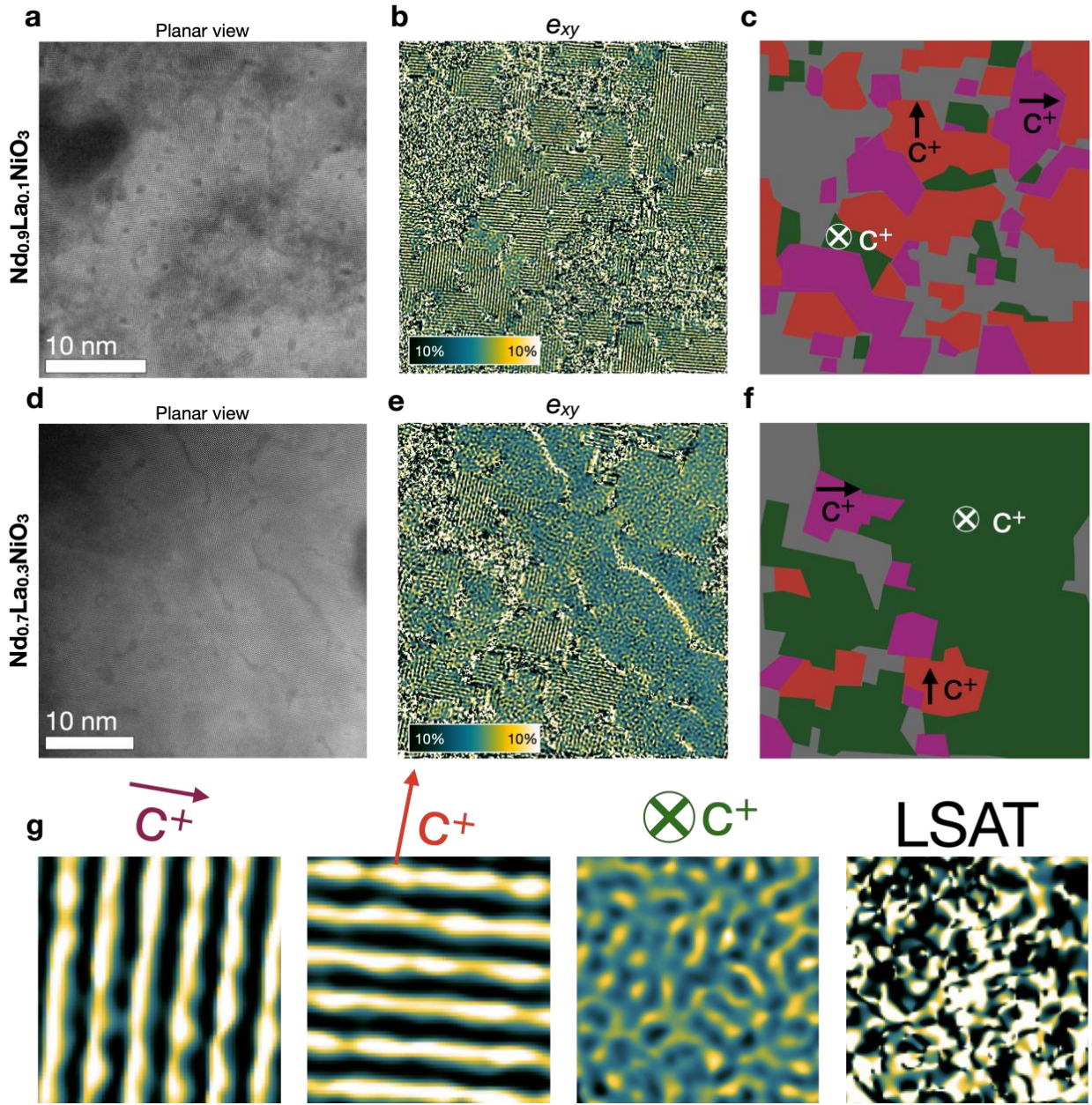

**Figure 4.** Plan-view HAADF image acquired in (a) a $Nd_{0.9}La_{0.1}NiO_3$ and in (d) a $Nd_{0.7}La_{0.3}NiO_3$ solid-solution thin film grown on LSAT. Their associated antipolar motion maps are presented in (b) and (e) respectively. (c), (f): orthorhombic domain distribution maps obtained from the antipolar motion maps displayed in (b) and (f). Each contrast feature, as all shown in (g), is indicated with a different color. The grey areas correspond to LSAT regions.

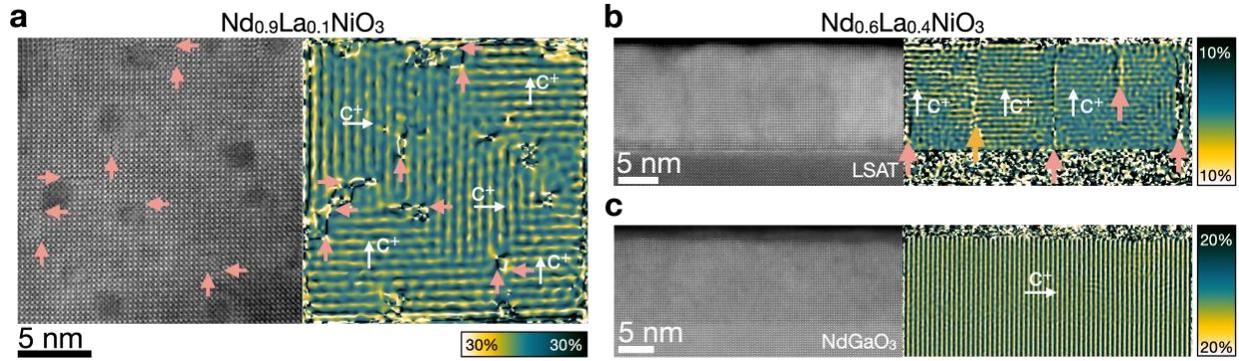

**Figure 5.** (a) Amplified view of a central region of the HAADF plan-view image and associated antipolar motion map displayed in Figure 4 (a), which was acquired from a $Nd_{0.9}La_{0.1}NiO_3$ solid-solution thin film deposited on $(001)_{PC}$-oriented LSAT. The red arrows point to RPFs present in the film, which tend to be localized at the boundaries between domains. (b) Cross-sectional HAADF image and associated antipolar motion map acquired from a $Nd_{0.6}La_{0.4}NiO_3$ solid-solution thin film deposited on $(001)_{PC}$-oriented LSAT. The red arrows similarly point to RPFs defects, while the orange arrow points to a twin (or antiphase) boundary. (c) Cross-sectional HAADF image and associated antipolar motion map acquired from a $Nd_{0.6}La_{0.4}NiO_3$ solid-solution thin film deposited on $(110)_{ORT}$-oriented $NdGaO_3$. $c_{ORT}$ is oriented along the in-plane direction, evidencing that the orientation of the lattice rotates by 90º, despite the similar strain values imposed by LSAT and $NdGaO_3$.